\documentclass[twocolumn,showpacs,prl]{revtex4}

\usepackage{graphicx}%
\usepackage{dcolumn}
\usepackage{amsmath}
\newcommand{\be}{\begin{equation}}
\newcommand{\ii}{{\rm i}}
\newcommand{\ee}{\end{equation}}
\newcommand{\bey}{\begin{eqnarray}}
\newcommand{\eey}{\end{eqnarray}}
\newcommand{\bw}{\begin{widetext}}
\newcommand{\ew}{\end{widetext}}
\newcommand{\ww}{\widetilde}
\newcommand{\ov}{\overline}
\newcommand{\ra}{\rangle}
\newcommand{\la}{\langle}
\newcommand{\br}{ {\bf r} }
\newcommand{\bp}{ {\bf p} }

\begin{document}

 \title {
 Uniform Semiclassical Approach to Fidelity Decay in the Deep Lyapunov Regime
 }

 \author{Wen-ge Wang,$^{1,2}$ G.~Casati$^{3,4,1}$, Baowen Li$^{1}$, and T.~Prosen$^{5,1}$}

 \affiliation{
 $^1$Department of Physics, National University of Singapore, 117542, Republic of Singapore
 \\ $^{2}$Department of Physics, Southeast University, Nanjing 210096, P.R.~China
 \\ $^{3}$Center for Nonlinear and Complex Systems, Universit\`{a}
 degli Studi dell'Insubria and Istituto Nazionale per la Fisica della Materia,
 Unit\`{a} di Como, Via Valleggio 11, 22100 Como, Italy
 \\ $^4$Istituto Nazionale di Fisica Nucleare, Sezione di Milano, Via Celoria 16, 20133 Milano, Italy
 \\ $^{5}$Physics Department, Faculty of Mathematics and Physics, University of
 Ljubljana, Ljubljana, Slovenia
 }

 \date{\today}

\begin{abstract}
 We use the uniform semiclassical approximation in order to derive
 the fidelity decay in the regime of large perturbations.
 Numerical computations are presented which
 agree with our theoretical predictions.
 Moreover, our theory allows us to explain previous findings,
 such as the deviation from the Lyapunov decay rate in cases
 where the classical finite-time instability is nonuniform in phase space.
\end{abstract}


\maketitle



  The stability of quantum motion under a system's perturbation can be measured  by the so-called
 quantum Loschmidt echo or fidelity \cite{Peres84,nc-book,gc-book}.
 It is defined as the overlap $ M(t) = |m(t) |^2 $ of two states obtained by evolving the same
 initial state $|\Psi_0\ra $ under two slightly different Hamiltonians:
 \be
 m(t) = \la \Psi_0|{\rm exp}(\ii Ht/ \hbar ) {\rm exp}(-\ii H_0t / \hbar) |\Psi_0 \ra .
 \label{mat} \ee
 Here $H_0$ is the Hamiltonian of a classically chaotic system and
 $ H=H_0 + \epsilon V $ is the perturbed Hamiltonian, with $\epsilon$ a small quantity and $V$
 a generic perturbing potential.
 This quantity can also be seen as a measure of the accuracy to which an initial quantum state
 can be recovered by inverting, at time $t$, the dynamics with the perturbed Hamiltonian $H$.

 This quantity has attracted much attention recently, mainly in relation to the field of quantum
 computation and in connection to the corresponding classical motion
 \cite{JP01,JAB02,CLMPV02,WC02,Prosen02,CT02,BC02,STB03,VH03,WCL04}.
 Focusing on systems with chaotic classical limit,
 one may identify,  by increasing the perturbation strength,
 three different regimes of fidelity decay:
 (i) The perturbative regime, in which the fidelity has a Gaussian decay.
 (ii) The Fermi-golden-rule  regime, with an exponential decay of  fidelity,
 $M(t) \propto \exp(-\Gamma t)$. Here the decay rate $\Gamma $ is the
 half-width of the local spectral density of states (LDOS) \cite{JAB02},
 which can also be calculated semiclassically \cite{CT02}.
 (iii) The Lyapunov regime, in which $M(t) \propto \exp(-\lambda t)$,
 with $\lambda$ being the (maximum) Lyapunov exponent of the underlying classical dynamics \cite{JP01}.

 However, the above picture remains unsatisfactory.
 This is particularly the case in the deep Lyapunov regime with $\sigma  \gg 1$,
 where $\sigma = \epsilon/\hbar$ 
 is the parameter characterizing the strength of quantum perturbation,
 and $\hbar $ the (effective) Planck constant.
 In  systems with nonconstant finite-time Lyapunov exponent (which is the typical situation),
 fidelity decays with a rate different from $\lambda $ \cite{STB03}.
 Indeed, a semiclassical analysis \cite{STB03} leads to an exponential
 decay of fidelity with a rate $\lambda_1 < \lambda $.
 The relation between this semiclassical treatment
 and that along the lines of Refs.~\cite{JP01,CT02,CLMPV02,VH03,WCL04} is unclear.
 Moreover, an extremely fast, super-exponential decay of fidelity has been found
 within a quite short initial time for initial Gaussian wave packets \cite{STB03}.
 In view of the importance of fidelity for the characterization
 of the stability of quantum motion under a system's perturbation,
 it is necessary to provide a clear theoretical understanding
 of its behavior and, in particular, to account for the seemingly
 disconnected and sometimes contradictory results.

 In this paper, we focus on the behavior of fidelity for $\sigma \gg 1$ and
 we treat this problem in full generality.
 We derive the general semiclassical
 formula which correctly reproduces the two limiting cases
 of $\exp(-\lambda t)$ and $\exp(-\lambda_1 t)$ decays.
 We also show that under certain conditions the exponential rate of
 fidelity decay can be equal to {\em twice} the classical Lyapunov exponent.

 Our starting point is the semiclassical approximation to the fidelity
 for an initial Gaussian wave packet given in \cite{VH03},
  \bey \nonumber  m_{\rm sc}(t)  \simeq \left (  \xi ^2 / \pi \hbar^2  \right )^{d/2}
 \hspace{3cm}
 \\ \times  \int  d{\bp_0}
 \exp\left [ \ii \Delta S({\bp_0}, \ww \br_0 ; t)/ {\hbar }
 - (\bp_0 - \ww \bp_0 )^2  / (\hbar / \xi  )^2   \right ],
 \label{mt-gauss-p0-1st} \eey
 where $\Delta S(\bp_0 , \ww \br_0 ; t)$ is the action difference
 along the two nearby trajectories starting at $(\bp_0 , \ww \br_0 )$ in the two systems $H$ and $H_0$,
 \be \Delta S(\bp_0 , \ww \br_0 ; t) \simeq \epsilon  \int_0^t dt' V[{\bf r'}(t')] \label{DS} \ee
 with $V$ evaluated along the trajectory in the $H_0$ system.
 The initial Gaussian wave packet, centered at ($\ww \br _0,\ww {\bf p}_0$),  is
 \be \label{Gauss-wp} \psi_0 (\br_0 ) = \left (  {\pi \xi ^2} \right )^{-d/4}
 {\rm exp} \left [ i \ww \bp_0 \cdot \br_0 / {\hbar}
 - (\br_0 - \ww \br_0 )^2/ (2 \xi ^2) \right ].  \ee
 For simplicity, we will consider here kicked systems with $d=1$
 and set the domains of $r$ and $p$ to be $[0,2\pi)$.
 The effective Planck constant is taken as $\hbar = 2\pi /N$, where $N$ is the dimension of the Hilbert space.

 The main feature of $\Delta S$ as a function of $p_0$ is
 its oscillations, the number of which increases exponentially with time $t$.
 Indeed, the variance of
 $\Delta S$ increases linearly with $t$ \cite{CT02}, while
 the slope of $\Delta S/ \epsilon $, denoted by $k_p$,
 \be k_p = \frac{1}{\epsilon } \frac{ \partial  \Delta S (p_0,r_0 ;t)}{\partial p_0}
 \simeq  \int_0^t dt' \frac{\partial V}{\partial r'}
 \frac{\partial r'(t')}{\partial p_0}, \label{partial-ds} \ee
 increases on average exponentially with $t$,
 due to the factor $\partial r' / \partial p_0 $.

 Let us first discuss the fidelity for a single initial state.
 Neglecting a quite short initial time,
 the main contribution to the right hand side of
 Eq.~(\ref{mt-gauss-p0-1st}) comes from the integration over the region $[\ww p_0 - w_p,\ww p_0 + w_p]$,
 where $w_p = \hbar / \xi  $ is the width of the initial Gaussian in the $p_0$ space.
 Let us define a time scale $\tau $ such that at $t = \tau $,
 $\Delta S(p_0,\ww r_0;t)$ completes one full oscillation period as
 $p_0$ runs over $[\ww p_0 - w_p,\ww p_0 + w_p]$.
 In a system possessing a constant local Lyapunov exponent $\lambda $,
 the number of  oscillations of $\Delta S$ increases exponentially as $c_0 e^{\lambda t}$
 and we obtain the average estimate for $\tau $,
 \be \ov \tau  \approx \frac{1}{\lambda } \ln ( \pi /c_0 w_p ) . \ee
 Taking, e.g., $\xi = \sqrt{\hbar }$, one can clearly see that $\ov \tau$
 is of the order of the Ehrenfest time $\frac{1}{\lambda } \ln \hbar ^{-1}$.
 In the general case of systems with fluctuation in the finite-time Lyapunov exponent,
 the number of the oscillations of $\Delta S$ increases as $e^{\Lambda (t) t}$,
 with some time-dependent rate $\Lambda (t) < \lambda $.

 We consider first the behavior of fidelity for $t<\tau $
 and denote with $\ww k_p $ the value of $k_p$ in the center $(\ww p_0, \ww r_0)$ of the initial Gaussian.
 For such times, the phase $\Delta S/ \hbar $, on the right hand side of Eq.~(\ref{mt-gauss-p0-1st}),
 as a function of $p_0$, can usually be approximated by a straight line with a slope $\sigma \ww k_p$,
 within the region $p_0 \in [\ww p_0 - w_p,\ww p_0 + w_p]$.
 Due to both the fast increasing of $|\ww k_p|$ with time
 and the large $\sigma $ value,
 one has $|\sigma \ww k_{p}| \gg \pi / w_p $ for most initial states and,
 as a result, the change of the phase $(\Delta S /\hbar )$ within the interval $p_0 \in [\ww p_0 - w_p,\ww p_0 + w_p]$
 is much larger than $2\pi $.
 Note that the largest slope of the term $(p_0 - \ww p_0 )^2 / w_p^2$ within this
 interval of $p_0$ is $2/w_p$, which is much smaller than $|\sigma \ww k_p|$.
 The right hand side of  Eq.~(\ref{mt-gauss-p0-1st}) can now be calculated approximately
 within the interval $p_0 \in [\ww p_0 - w_p,\ww p_0 + w_p]$ and gives
 \be M_{\rm sc}(t) \propto 1/ (\sigma \ww k_{p})^2. \label{M-skp} \ee

 \begin{figure}
 \includegraphics[width=\columnwidth]{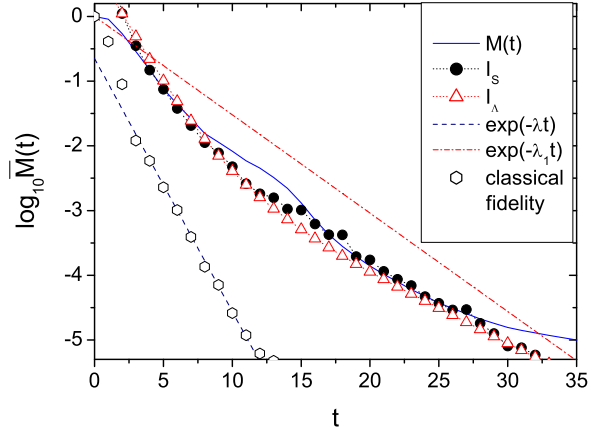}
 \vspace{-1.1cm}
 \caption{
 Decay of averaged fidelity in the map (\ref{Liverani}),
 with $K =1$, $\eta =0.987$.
 $I_s$ and $I_{\Lambda}$ are the theoretical predictions (\ref{Mt-It})
 and (\ref{It-3}), respectively.
 It is seen that after a short initial time,
 both $I_s(t)$ and $I_{\Lambda }(t)$ are close to the exact fidelity $\ov M(t)$ (until saturation is reached).
 For comparison, the decay $e^{-\lambda_1 t}$ is shown.
 We also plot the average classical fidelity, which was calculated by taking initial
 points within circles with radius $\sqrt{\hbar }$  in the phase space  \cite{BC02,VP03}.
 For this map, $\lambda \approx 0.9$ and $\lambda_1 \approx 0.35$.
 Here and in the following figures,
 $\sigma =100$, $N=131\ 072$, $\xi  = \sqrt{\hbar }$,
 and averages are performed over 2000 initial Gaussian packets.
 } \label{fig-Mt-sk-0987}
 \end{figure}

 For times $t> \tau $, or when $|\ww k_{p}|$ is small enough for $t < \tau$,
 the stationary phase approximation can be used in calculating $m_{\rm sc}(t)$
 in Eq.~(\ref{mt-gauss-p0-1st}).
 If we denote by $\alpha $ the stationary points and by $p_{0\alpha }$
 the momenta at which $k_p=0$,
 we have $m_{\rm sc}(t) \simeq \sum_{\alpha } m_{\alpha }(t)$, where
 \bey m_{\alpha }(t) = \frac{\sqrt{ 2 i \hbar }}{w_p }
 \frac{ {\rm exp} \left [ \frac i{\hbar} \Delta S( p_{0\alpha },\ww r_0 ; t)
 - (p_{0\alpha } - \ww p_0 )^2/w_p^2 \right ] }
 {\sqrt{ | \Delta {S_{\alpha }^{''}} | }}, \nonumber
 \\ \text{with} \ \ \ \  \Delta {S_{\alpha }^{''}} = \left .
 \frac{ \partial^2 \Delta S(p_0,\ww r_0;t) }{ \partial p_0^2 }
 \right |_{p_0= p_{0\alpha }}. \hspace{0.3cm}  \label{ma-st} \eey

 Next we turn to the behavior of average fidelity
 and first consider the long time decay, namely, $t> \tau $.
 Due to the large  $\sigma $ value and to the classically chaotic motion,
 the phase $\Delta S(p_{0\alpha },\ww r_0;t) / \hbar $ in Eq.~(\ref{ma-st})
 can be regarded as random with respect to $\alpha $ and $\ww r_0$.
 Then, the averaged fidelity $\overline M(t)$, with average taken over
 $\ww r_0$ and $\ww p_0 $, can be approximated by its diagonal part \cite{WCL04},
 \be \ov M(t) \simeq \ov { \sum_{\alpha } |m_{\alpha }(t)|^2 }
 \simeq \frac {\xi  }{(2\pi)^{3/2} } \int_0^{2\pi} d \ww r_0 \sum_{\alpha }
 \frac 1{ \left | \Delta {S_{\alpha }^{''}} \right |} . \label{M-st} \ee

 \begin{figure}
 \includegraphics[width=\columnwidth]{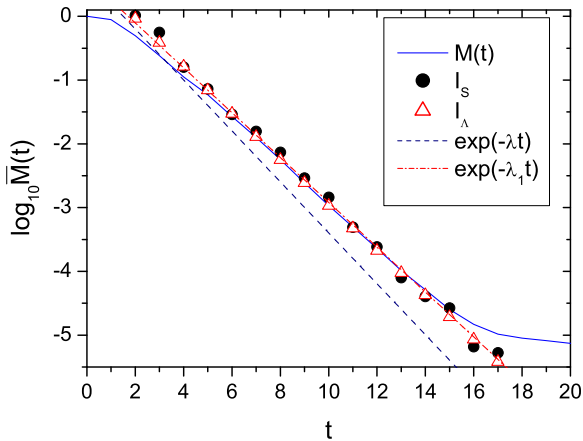}
 \vspace{-1.2cm}
 \caption{
 Same as Fig.~\ref{fig-Mt-sk-0987}, but for $\eta =0.85$, for which
 $\lambda  \approx 0.92$ and $\lambda_1 \approx 0.81$.
  } \label{fig-Mt-sk-085}
 \end{figure}

    The right hand side of Eq.~(\ref{M-st}) can be expressed as an integration of
 $1/|k_p |$.
 For this, we introduce ${\cal A}_{\alpha }$ to denote the region
 $[p_{0\alpha }^-,p_{0\alpha }-\delta ] \bigcup [p_{0\alpha }+ \delta ,p_{0\alpha}^+]$,
 where $p_{0\alpha}^{-}= ({p}_{0\alpha } +{p}_{0,\alpha -1})/2$,
 $p_{0\alpha}^{+}= ({p}_{0\alpha } +{p}_{0,\alpha +1})/2$,
 and where $\delta $ is a small quantity.
 In the neighborhood of $p_{0\alpha }$,
 $k_p$ satisfies $\epsilon k_p  \simeq \Delta {S_{\alpha }^{''}} \cdot (p_0-p_{0\alpha})$.
 For small enough $\delta$, we have
 \be \int_{ {\cal A}_{\alpha } } dp_0 \frac 1{ | \epsilon k_p |}
 \simeq -\frac{ 2 \ln \delta }{ \left | \Delta {S_{\alpha }^{''}} \right | }. \label{int-ds1} \ee
 Substituting the expression of $|\Delta S_{\alpha }''|$ obtained from
 Eq.~(\ref{int-ds1}) into Eq.~(\ref{M-st}), we have
 \be \ov M(t) \approx
 \frac{\xi  }{(2\pi )^{3/2} (-2\ln \delta ) \epsilon}  \int_0^{2\pi } d
 \ww r_0 \int_{ {\cal P}_{\delta }} dp_0 \frac 1{ | k_p |}, \label{Mt-delta} \ee
 where ${\cal P}_{\delta } := \bigcup_{\alpha } {\cal A}_{\alpha }$.

   Since the value of $\delta $ is irrelevant for the decay rate, we may write
 \be \label{Mt-It}  \ov M(t) \propto  I_s(t) := \int d \ww r_0 \int_{ {\cal P}_{\delta }} dp_0
 \frac 1{ | k_p |}. \ee
 An accurate numerical evaluation of $I_s(t)$ is
 not easy since one must find out all stationary points $\alpha $ for each
 value of $\ww r_0$.
 An approximate numerical result can be obtained by using the Monte Carlo method
 in which, in order to perform the integral (\ref{Mt-It}) over the region ${\cal P}_{\delta }$,
 i.e., with the neighborhoods of stationary points excluded,
 we neglect the small set of points
 that have the smallest values of $|k_p|$.

 Actually, one can make a further approximation by using the following arguments.
 The main contribution to the integral in Eq.~(\ref{Mt-delta})
 comes from small values of $|k_p|$ in the region ${\cal P}_{\delta }$.
 For $p_0 \in {\cal P}_{\delta }$ close to a stationary point $p_{0\alpha }$,
 $k_p$ in Eq. (\ref{partial-ds}) can be approximated by
 \be k_p \approx \int_0^t dt' \left [
 \frac{\partial^2 V}{\partial r'^2 } \left ( \frac{\partial  r'}{\partial p_0} \right )^2
 + \frac{\partial V}{\partial r' } \frac{\partial^2  r'}{\partial p_0^2}
 \right ] (p_0-p_{0\alpha }) . \label{kp-2p0} \ee
 Due to exponential divergence of neighboring trajectories in phase space,
 the main contribution to the right hand side of Eq.~(\ref{kp-2p0})
 comes from times $t' \approx t$.
 The time evolution of the quantity inside the bracket in Eq.~(\ref{kp-2p0}) is given by
 the dynamics of the system described by $H_0$.
 On average it increases as $ [\delta x(t)/ \delta x(0)]^2$, where $\delta (x)$ denotes
 distance in phase space.
 With increasing time, the number of the stationary points of $\Delta S$ increases exponentially,
 roughly in the same way as  $ \delta x(t)/ \delta x(0)$,
 since the oscillation of $\Delta S$ is mainly induced by local instability
 of trajectories.
 Then, substituting Eq.~(\ref{kp-2p0}) into Eq.~(\ref{Mt-delta}), we have
 $ \ov M(t) \propto \ov {|\delta x(t)/ \delta x(0)|^{-1}} $,
 which can be written as
 \bey \nonumber   \ov M(t) & \propto & I_{\Lambda}(t) = \exp [{-\Lambda_1 (t) t}], \ \ \ \ \  \text{with}
 \\ \Lambda_1 (t) & = & - \frac 1t  \lim_{\delta x(0) \to 0} \ln
 \overline{ \left |\frac{ \delta x(t)}{ \delta x(0)} \right |^{-1} }.  \label{It-3} \eey
 In systems with constant local Lyapunov exponents,
 Eq.~(\ref{It-3}) reduces to the usual Lyapunov decay with $\Lambda_1 (t) =\lambda $.
 On the other hand, when fluctuations in local Lyapunov exponent cannot be neglected,
 $I_{\Lambda }(t)$ coincides with the $e^{-\lambda_1t}$ decay in Ref.~\cite{STB03}
 with $\lambda _1 = \lim_{t \to \infty } \Lambda_1 (t) < \lambda $,
 {\it only} in the limit $t \to \infty $.
 Therefore, the actual decay, which can be observed in finite times,
 can be considerably different from the $e^{-\lambda_1 t}$ decay.

 For times $t< \tau $,
 the main contribution to the averaged fidelity $\ov M(t)$ comes, in fact, from
 initial states with $\ww k_p$ close to zero.
 When $\Delta S$ is not too flat,
 one can still use the stationary phase approximation for these initial states.
 Hence, also for $t< \tau $ we obtain the same expressions as in Eqs.~(\ref{Mt-It})
 and (\ref{It-3}) for the decay of averaged fidelity.

 \begin{figure}
 \includegraphics[width=\columnwidth]{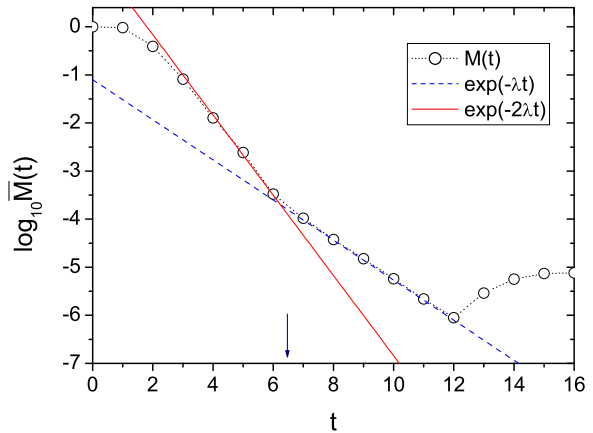}
 \vspace{-1.2cm}
 \caption{
 Decay of averaged fidelity in the sawtooth map ($\eta =0$) with $K=1$
 and $i=3$ in Eqs.~(\ref{Liverani}) and (\ref{Vi}),
 showing $e^{-2\lambda t}$ decay followed by the Lyapunov decay
 ($\lambda =0.96$). The arrow indicates the theoretical estimate of the crossover time $\ov\tau$.
 } \label{fig-Mt-st-v3}
 \end{figure}

 We have tested the above predictions by considering the map
 \begin{eqnarray} \nonumber
 p_{n+1} &=& p_n + K [(r_n-\pi ) + \eta  \sin r_n ] \pmod{2\pi},\\
 r_{n+1} &=& r_n +p_{n+1} \hspace{2.5cm} \pmod{2\pi}, \label{Liverani}
\end{eqnarray}
 with two parameters $K,\eta \in [0,1]$.
 For $K >0$ and $\eta = 0$, this is the piece-wise linear sawtooth map \cite{BC02},
 which is hyperbolic with constant local (finite time) Lyapunov exponent.
 For the particular case  $K=1$, the map reduces to the perturbed cat
 map, which is known to be Anosov for $0 < \eta < 1$ (having
 non-constant $\lambda$), whereas for $\eta=1$ it acquires a
 marginally stable (parabolic) fixed point.
 This map is quantized in a Hilbert space of dimension $N$.
 The one period quantum evolution is given by the Floquet operator,
 $ \label{U} U = \exp [-\ii {\hat p}^2 / (2 \hbar )] \exp [-\ii U( {\hat r}) / \hbar ] $,
 with $U(r)= -K[ (r-\pi)^2 /2 - \eta \cos r]$.
 In order to compute fidelity, we choose to perturb the parameter $K \to K + \epsilon $.
 Figure \ref{fig-Mt-sk-0987} shows that numerical data accurately fit our theoretical
 predictions in Eqs.~(\ref{Mt-It}) and (\ref{It-3}).
 In Fig.~\ref{fig-Mt-sk-085}, it is seen that
 with decreasing $\eta $, since the values of $\lambda$ and $\lambda_1 $ become closer,
 our predictions approach that of Ref.~\cite{STB03}.
 At $\eta =0$, the classical map has a constant local Lyapunov exponent and
 the standard Lyapunov decay is recovered.

 In the above discussion of the average fidelity,
 the existence of stationary phase is assumed.
 It may happen, in some circumstance, e.g., with some
 special perturbation, that there is no stationary phase for $\Delta S $.
 In this case it turns out that
 a decay with a rate of double Lyapunov exponent may appear for $t<\tau $,
 when the classical system has a constant local Lyapunov exponent.
 Indeed, for $t < \tau  $, the main contribution to the averaged fidelity $\ov M(t)$
 comes from initial states for which the values of $|\ww k_p|$
 are close to local minimum of $|k_p |$.
 When the values of local minimum of $|k_p|$ are large enough,
 the decay of the fidelity is given by Eq.~(\ref{M-skp}).
 Then, since $|k_p|$ increases on average as $e^{\lambda t}$,
 the averaged fidelity has a double-Lyapunov-exponent decay,
 \be \label{ovM-2} \ov M(t) \propto e^{-2\lambda t},
 \hspace{0.5cm} t < \tau . \ee
 For $t>\tau $,
 one can use arguments given in
 Ref.~\cite{WCL04}, showing that $\ov M(t)$ still follows the standard Lyapunov decay,
 $\ov M(t) \propto e^{-\lambda t}$.

 \begin{figure}
 \includegraphics[width=\columnwidth]{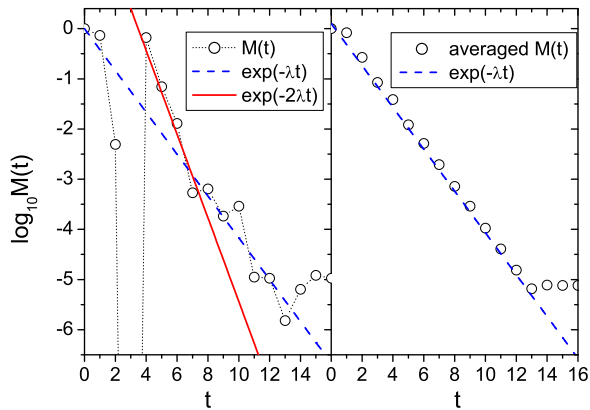}
 \vspace{-1.2cm}
 \caption{
 Fidelity decay in the sawtooth map ($\eta =0$) with $K=1$ and $i=2$.
 Left panel: $M(t)$ of a single initial Gaussian,
 showing large fluctuation at $t<4$, approximate $e^{-2\lambda t}$
 decay within $4 \le t \le 7$, and approximate Lyapunov decay
 at $t \ge 8$ (before saturation).
 Right panel: averaged fidelity $\ov M(t)$, showing the Lyapunov decay.
 } \label{fig-Mt-st-v2}
 \end{figure}

 Finally, in systems possessing stationary phase in $\Delta S$
 and constant local Lyapunov exponents,
 although the averaged fidelity has Lyapunov decay,
 a double-Lyapunov-exponent decay $e^{-2\lambda t}$ may appear for $t< \tau $,
 for the fidelity of those {\it single} initial states,
 for which $|k_p|$ happens to increase exponentially as $e^{\lambda t}$
 [see Eq.~(\ref{M-skp})].

 In order to check the above predictions, we consider the sawtooth map ($\eta =0$)
 which has a constant local Lyapunov exponent,
 $\lambda = \ln ( \{ 2+ K + [ ( 2+K)^2 -4 ]^{1/2} \} /2 )$.
 We consider here the following perturbed map
 \bey \nonumber
 p_{n+1} &=& p_n + K (r_n-\pi ) + \epsilon i {\cal N}_i (r_n -\pi )^{i-1},
 \ \ \ i=2,3 ,\\
 r_{n+1} &=& r_n +p_{n+1}, \label{Vi}
 \eey
 with ${\cal N}_2 = 1/2$ and $ {\cal N}_3 = \sqrt {1.4} /3 \pi $.
 These two values ${\cal N}_i$ give the same decay rate in the Fermi-golden-rule regime.
 However, while for $i=2$ stationary phase of $\Delta S$ exists,
 in the case $i=3$ there is no stationary phase in $\Delta S$ vs $p_0$.
 In the latter case, as shown in Fig.~\ref{fig-Mt-st-v3}, the average fidelity
 has an initial double-Lyapunov-exponent decay followed by the standard Lyapunov decay,
 as predicted by the theory.
 The crossover of the two decays is
 in agreement with the theoretical estimate  $\ov \tau  \approx 6.5$.
 Figure \ref{fig-Mt-st-v2} (left panel) shows instead that a
 double-Lyapunov-exponent decay may appear for the fidelity of some particular single
 initial state,
 while the average fidelity  has the Lyapunov decay (right panel).

 In summary, we have derived general semiclassical
 expressions for the fidelity decay, at strong perturbations,
 which reproduce, as two particular limiting cases, previous results
 leading to the Lyapunov decay and to the $e^{-\lambda_1 t}$ decay.
 In particular we have discussed the relevance  of
 fluctuations in the finite-time Lyapunov exponent and we have shown that fidelity decay
 depends on the strength of such fluctuations in the Lyapunov regime.

 This work was supported in part by the Academic Research Fund of the National University of Singapore
 and the Temasek Young Investigator Award (B.L.) of DSTA Singapore under Project Agreement No.~POD0410553.
 Support was also given by the EC RTN contract No.~HPRN-CT-2000-0156, the NSA and ARDA under ARO contracts Nos.
 DAAD19-02-1-0086, the project EDIQIP of the IST-FET programme of
 the EC, the PRIN-2002 ``Fault tolerance,
 control and stability in quantum information processing'',
 and Natural Science Foundation of China Grant No.10275011.

 \end{document}